# Shape tailoring to enhance and tune the properties of graphene nanomechanical resonators

*David Miller, Benjamín Alemán*

Department of Physics, University of Oregon, Eugene, Oregon 97403

Material Science Institute, University of Oregon, Eugene, Oregon 97403

Center for Optical, Molecular, and Quantum Science, University of Oregon, Eugene, Oregon 97403

**Abstract**

**The shape of a nanomechanical resonator profoundly affects its mechanical properties and determines its suitability for various applications, such as ultra-sensitive mass and force detection. Despite the promise of two-dimensional nanomechanical systems in such applications, full control over the shape of suspended two-dimensional materials, such as graphene, has not been achieved. We present an effective, single-step method to shape pre-suspended graphene into nanomechanical resonators with arbitrary geometries leading to enhanced properties in comparison to conventional drumheads. Our technique employs focused ion beam milling and achieves feature sizes ranging from a few tens of nanometers to several microns, while obtaining near perfect yield. We compare the mechanical properties of the shaped devices to unmodified drumheads, and find that low-tension, singly-clamped graphene cantilevers display a 20-fold increase in the mechanical quality factor (Q) with a factor 100 reduction in the mechanical damping. Importantly, we achieve these results while simultaneously removing mass, which enables state-of-the-art force sensitivity for a graphene mechanical resonator at room temperature. Our approach opens up a unique, currently inaccessible regime in graphene nanomechanics, one characterized by low strain, low frequency, small mass, and high Q, and facilitates tailoring of non-linearity and damping in mechanical structures composed of graphene and other 2D crystals.**

**Introduction**

Nanomechanical resonators, such as freely vibrating nanometer-scale beams and membranes, have enabled ultrasensitive physical measurements at the level of single atom mass[1] and single electron charge[2] as well as the exploration of quantum mechanics in macroscopic mechanical systems[3]. Among the most promising applications of nanomechanical systems is the ability to detect extremely small forces, such as those that arise from chemical or biological processes[4] or from electronic or nuclear spins[5], which is ultimately limited by thermal fluctuations due to mechanical damping. The minimum detectable force[6] for a mechanical resonator at a given temperature is directly related to the coefficient of mechanical damping as $dF_{min} \propto \sqrt{b}$, where the damping coefficient, b, is related to the resonator mass, mechanical resonance frequency and quality factor through $b \propto m_{eff} f_0 / Q$. Thus, the ideal force sensor would have low mass, relatively low tension, and a high quality factor.

Low-dimensional materials such as nanotubes and two-dimensional crystals, including graphene, have begun to see wide use as nanomechanical systems because of their inherently small mass and strong interactions with their environment[7–10]. Graphene is exceptionally well-suited for nanomechanical systems because it also offers high intrinsic stiffness, strength[11], and amenability to strain tuning[12]. Additionally, fabrication of large-scale arrays of graphene drumhead resonators is well developed[13] and drumheads are frequently used in nanomechanical experiments. However, although various techniques have been used to increase the quality factor in graphene drumheads, such as using of few-layer reduced graphene oxide membranes[14], or larger area graphene drumheads[15], they also add significant mass, leaving the force sensitivity unchanged. Thus, new approaches are required to realize graphene mechanical resonators that both have high quality factors and the ultra-low mass.

Tailoring the geometric shape and clamping of nanomechanical resonators is a promising alternative to achieve reductions to the mass while also lowering tension and damping[16,17]. Although such geometric tuning of graphene is still in its infancy, the few studies that have explored geometric effects indicate that shape and size has a large role in the mechanical properties of graphene resonators. For example, low tension H-shaped graphene suspended structures[18] were found to display order-of-magnitude increases to the mechanical Q along with a significantly reduced damping coefficient. In contrast, doubly-clamped beams[19] show quality factors and mass similar to graphene drums, indicating the need for more detailed studies to elucidate the role of geometry and tension on the mechanical properties of graphene resonators. However, the arbitrary shaping of suspended graphene remains elusive, which is in part due to current fabrication approaches, so many potentially compelling device geometries, even as simple as a singly clamped cantilever, have yet to be fully explored.

Fabrication of arbitrarily patterned graphene mechanical structures via resist-based lithography[19,20] and planar processing has not been achieved. This is partly due to the cumbersome, multistep nature of clamping and suspending such devices, which involves multiple lithography steps, thin film depositions, dry and wet etching, and critical point drying. In many cases, the etching chemistry needed to define the mechanical clamp is incompatible with graphene, which precludes the approach altogether. An alternative patterning approach, one that circumvents challenges seen in traditional lithography, has emerged that employs Focused Ion Beam (FIB) milling[21] of free-standing graphene. This approach has been used to pattern graphene into diffraction gratings[22], nanopores[23], and nanowires[24]. The FIB technique has seen little use as a method to pattern single-layer graphene nanomechanical systems and has presently only been used to fabricate low-aspect ratio cantilevers[25] with no associated improvements to the Q. Thus, the viability of FIB milling as a general approach to achieve arbitrarily shaped graphene mechanical resonators remains an open question. Furthermore, because the geometric parameter space of graphene nanomechanical resonators is largely unexplored, it is unknown which shapes or clamping configurations possess less mechanical damping.

In this letter, we demonstrate that FIB milling is an effective tool to shape free-standing graphene membranes into a wide variety of two-dimensional geometries, with device features ranging in size from several tens of nanometers to a few micrometers. Many of these structures, such as crosses, triangular cantilevers, and tethered cantilevers, have not been previously observed in a suspended two-dimensional material. Furthermore, we employ optical techniques to actuate and detect the mechanical motion of the graphene structures in order to characterize their mechanical properties, such as the Q, resonance frequencies, and force sensitivities. We compare unmodified drumheads to the FIB milled structures and identify that singly-clamped graphene devices can display order-of-magnitude enhancements to the quality factor while also reducing mass, making them an ideal candidate for graphene force sensors. We also demonstrate that shape can be used to introduce mechanical nonlinear behavior and also stabilize the frequency of devices under optical probing, showing the broad generality of nanomechanical properties that may be tuned through geometric shape.

**Fabrication of graphene nanomechanical resonators with arbitrary geometry**

The starting template for the shaped graphene devices is a graphene sheet suspended over a pre-patterned circular hole, forming a freely suspended graphene mechanical resonator with uniform edge clamping (*i.e.* a circular drumhead). We used commercially available single-layer graphene on holey silicon nitride grids (Ted Pella Part# 21712) for device templates[21,22]. Each grid contains a periodic array of several thousand individual circular drumheads, each with a diameter of 2.5 μm. To characterize the quality of the graphene prior to milling, we used transmission electron microscopy (TEM) and Raman microscopy. We observe some degree of surface contamination under TEM and SEM, which is an unavoidable byproduct when transferring CVD graphene using standard polymer-based techniques (Figure 1 and Figure S1). The Raman spectrum typical of low-defect, annealed monolayer graphene (Figure S2) that is relatively free of defects [26]. We also use selected area electron diffraction (SAED) to confirm the crystalline, single-layer nature of the graphene (Figure S3).

Graphene resonators were shaped by irradiative milling of the suspended graphene membrane template with a focused ion beam or FIB. The "positive-tone" FIB milling process sputters material from specified regions of the membrane to obtain the desired device geometry. Milling was accomplished with a commercial gallium FIB (FEI Helios 600i Ga+ SEM-FIB) operated in vacuum at 30 kV and with 1.1 pA ion currents to minimize damage due to the spread of the ion beam. Typical ion doses required to mill through the graphene were 8.5-17 pC/μm$^2$, corresponding to 0.06-0.12 μm$^2$/s milling rates. Prior to fabricating devices, a brief snapshot image was taken with the FIB to orient the milling patterns. Snapshots were taken of drumheads as well. During the snapshot, we apply an ion dose of ~.0007 pC/μm$^2$, which is 10000 times less than the dose required to mill graphene. Examples of the FIB milled geometries are shown in Figure 1 and Figure S4. To demonstrate the flexibility and robustness of our technique, we fabricated similar devices from graphene suspended over cavities (Figure S4).

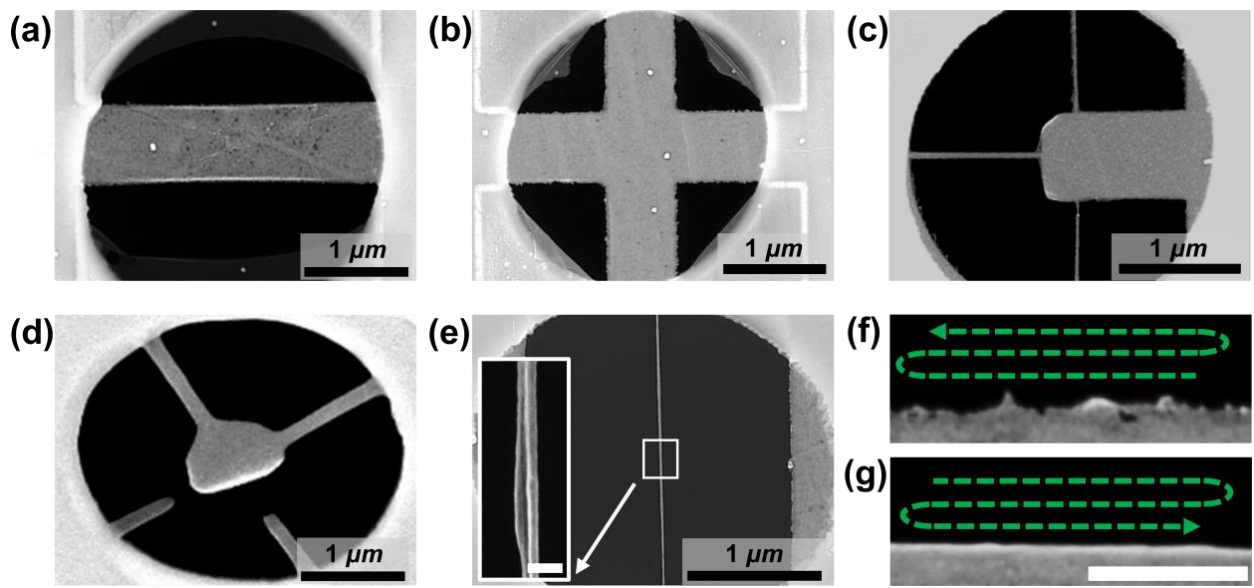

**Figure 1: Gallery of SEM images of graphene devices fabricated with focused ion beam milling.** (a) Graphene doubly-clamped beam with 600 nm width. (b) Cross with 600 nm bar widths. Peeled areas of graphene are visible around the edge of the circular hole. (c) Tethered cantilever with aspect ratio of 1.66 achieved through use of stabilizing tethers (d) Triangular cantilever with a 90 degree tether angle. (e) Graphene scroll with width ~25 nm spanning across the entire hole with a 100:1 aspect ratio. Rolling of the graphene is visible in the inset (Scale = 70 nm). (f) Edge of a graphene beam where the ion beam mills away from the device with local RMS roughness of 2.47 nm (g) Edge of a graphene beam where the ion beam mills towards the device with local RMS roughness of 0.23 nm.

We use four types of cuts to pattern the graphene. These cuts include a vector cut, where a single line is milled into the graphene with width determined by the Gaussian beam waist of the ion beam, a raster cut, where the beam passes over an area in many passes, and two types of single-pass directional raster cuts, shown in Figure 1(f-g), where the serpentine raster is either directed away from or towards the device. The type and order of cuts dictated possible device geometries. In some cases, using the incorrect sequence of these cuts led to device failure.

We monitored the entire cutting process using the non-destructive scanning electron microscope (SEM) imaging system before, during, and after fabrication (Figure 2 and Supporting Videos SV1-SV3). This allowed us to fabricate devices in regions with fewer particulate contaminates, holes, and folding (multilayer) defects, while also allowing us to determine successful cutting strategies for each of the device geometries. For instance, we could observe if a particular cut caused device failure through tearing or rupturing and subsequently adjust the cutting sequence or type accordingly. Post-fabrication SEM characterization generated maps of devices, which were used during optical characterization to locate and probe specific devices. We used the FIB patterning approach described above to generate a variety of device geometries. These include crosses, beams, two cantilever style geometries (Figure 1), coupled beams (Figure 2(b)), meshes, scrolls, and tethered trampolines (Figure S4). Many of these geometries have not been previously achieved in suspended graphene. This patterning technique achieved feature sizes as narrow as 10 nm, pitch resolution less than 100 nm, and length-to-width ratios as high as 250:1. We also generated edge-clamping configurations ranging from double-clamping (*e.g.* in simple beams) to 48 independent clamps (*e.g.* in trampolines), with clamp widths ranging from 10 nm to 1 μm.

Each device architecture required a particular, manually defined sequence of FIB cuts, which was largely determined by the need to manage tension or strain during device fabrication. Tensioned graphene, unlike many commonly FIB milled bulk materials[27], such as silicon or diamond, is susceptible to warping, tearing, and rupturing due to asymmetric strain that is introduced during FIB milling. An illustration of tension-driven failure in a simple beam device is shown in Figure 2(a). In this example, an initial raster cut removed graphene from the left half of the graphene drumhead, resulting in tension originating only from edge-clamping on the right half of the membrane. As milling proceeded on the right side of the drumhead, tension became concentrated near the center causing the device to stretch and then tear. We observed that larger milled regions led to a greater tension imbalance around small device features, limiting FIB milling to areas of less than ~500 nm in lateral dimensions when tension asymmetries were not managed and controlled. However, once the proper cutting sequence for a given geometry was established, fabrication yield was near 100%.

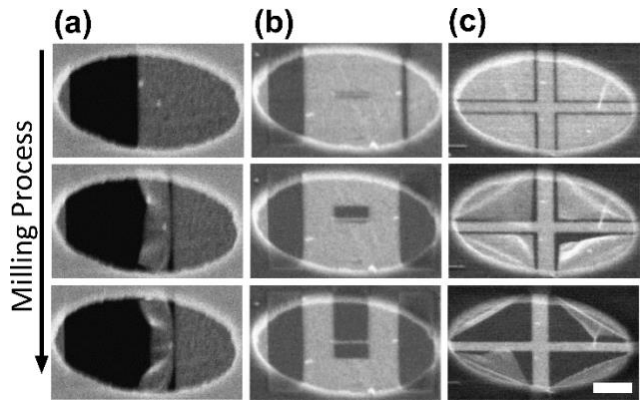

**Figure 2: *In-situ* SEM observation during FIB milling of graphene devices.** (a) Sequence of images showing a failure during fabrication of a doubly-clamped beam. (b) Successful fabrication of a coupled beam. The narrow central ribbon is protected from excess tension by two isolating vector cuts in the center of the structure. (c) One of several fabrication methods for a cross. Vector cuts are used to outline the cross shape. Then, a parallel raster peels the graphene away from the device**.** (Scale Bar = 500 nm)

We identified several methods to controllably relieve tension in the devices to avoid tension-driven failure. These include specifying the mill direction, specifying the order of particular cuts, and using single-pass or multiple pass milling.

Simple structures such as crosses and doubly-clamped beams could be shaped with high repeatability through several methods, including directional single-pass raster cuts or a vector cutting method shown in Figure 2(c). In the vector cutting method, a series of vector cuts are used to etch the outline of the shape into the graphene. The ion beam then rasters around the region, causing the graphene to peel away from the device.

The coupled beam geometry, consisting of two wide doubly clamped beams (500 nm wide, 2.5 μm long) coupled together through a narrow ribbon (50 nm wide, 500 nm long), required management of tension around the central ribbon, which was very sensitive to asymmetric tension. One successful milling sequence, shown in Figure 2(b), starts by defining the narrow ribbon vector cuts on both sides to isolate it from tension imparted during later milling. Then, a single pass raster on both sides of the drumheads leaves a single 1.5 μm by 2.5 μm beam. Finally, single pass raster cuts on either side of the thin ribbon to leave the freestanding, coupled beam geometry. This structure is the first example of coupled beams in graphene, which have been previously shown[28] to display complex non-linear dynamics and chaos in resonators fabricated from bulk materials.

We were able to fashion the graphene into nanoribbons with widths of 40 nm and lengths of 2.5 μm, which we achieved using a single pass directed raster towards the ribbon. The raster direction here was crucial, as outward raster cuts or multiple pass raster cuts frequently resulted in failure of the tether. In contrast, an inward raster severs the edge clamp first in order to relieve strain and thereby stabilize the ribbon as it forms. By reducing the ribbon width below ~40 nm, the ribbon spontaneously narrows and changes into a structure resembling a nanoscroll[22]. We achieved nanoscrolls with widths of 10-15 nm that spanned the entire 2.5 μm width of the drumhead template, yielding an aspect ratios as high as 250:1. The nanoscroll and nanoribbon structures were fashioned as stand-alone devices (Figure 1(e)) and also served as tethers in more complex structures such as tethered cantilevers (Figure 1(c)) and trampolines (Figure S4).

The raster direction relative to the edge of a device feature also affected the RMS roughness of the edge. A raster away from an edge with a single pass (Figure 1(f) and Figure S1) resulted in a local edge roughness of 2.47 nm. A raster towards an edge resulted in a smoother edge with an edge roughness of 0.23 nm (Figure 1(g)). Based on SEM, these smooth edges are likely due to scrolling similar to that evident in the device in Figure 1(d). Edge roughness can lower the thermal conductivity[29], reduce electron mobility[30], and increase damping[31] of graphene devices and reducing the edge roughness using FIB milling could be an effective route towards improving these characteristics.

FIB milling introduces some degree of defects and contamination when milling bulk materials or graphene[32]. We investigated these effects with Raman and TEM. Even at the relatively low ion doses used in this work, both the lightly dosed drumheads and the milled devices had Raman spectra consistent with increased disorder in the graphene (Figure S2). This is in accord with previous studies of FIB milled or otherwise patterned monolayer graphene[33–35]. We attribute this damage to deposition of amorphous carbon during SEM imaging [36] or FIB milling, to the FIB snapshot images taken to orient the milling, and to the FIB fabrication itself. We also expect the cut edges in the FIB milled devices to contribute significantly to the disorder in the observed Raman data[37]. To confirm that the fabricated devices are still crystalline, we perform SAED using TEM on the graphene before and after FIB irradiation, and we observe no difference in the diffraction patterns (Figure S3), so milled devices remain crystalline. Since all the devices studied in this work were exposed to a similar, relatively low amount of ion irradiation, we attribute the enhanced mechanical properties described below to the geometric shape rather than the FIB irradiation. Damage due to the FIB process could be reduced in future work through use of more localized etching processes, such as helium FIB milling[34] or water-assisted etching[35], or by a post-fabrication annealing step.

**Mechanical characteristics of shaped graphene devices**

Having used FIB milling to demonstrate robust and reproducible control over the geometric shape of suspended graphene mechanical structures, we now turn to discussing the mechanical properties of some of these structures. Our central data include amplitude and phase spectra obtained via Michelson interferometry[38] modeled with a driven damped harmonic oscillator to infer the Q, damping, mode frequencies ($f_n$), and corresponding amplitudes. We optically drove the mechanical resonators with an amplitude modulated 445 nm blue laser, with tunable power output, $P_0$. The power of the blue laser incident on the drumheads is given by $P=\frac{P_0}{2}(1+\cos(f \cdot t))$, which has an AC term, leading to photothermal actuation[13,15], as well as a DC term, leading to optical heating and increased strain in the devices[9]. A detailed diagram of the optical experiments is shown in Figure S5.

We first probed the amplitude response of drumhead resonators (Figure 3) to establish a baseline for comparisons with etched geometries. Although these drumheads were not ion milled, they were irradiated through the initial 'snapshot' of the devices. We measured eleven 2.5 μm diameter drumheads and found a center frequency $f_0$=21.54±4.79 MHz, Q=48.85±13.04, and a damping coefficient of b=2.7 pg/s. From $f_0$, we calculate a minimum possible strain of strain of $\epsilon \sim 1 \times 10^{-5}$ (Section S6), which is comparable to previous measurements of drumheads using graphene grown via chemical vapor deposition and transferred using sacrificial polymer layers[15].

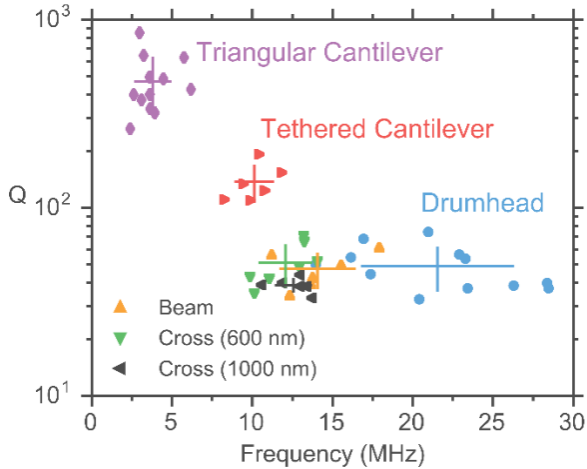

**Figure 3**: **Measured resonance frequencies and quality factors for FIB shaped devices.** Error bars indicate the standard deviation of $f_0$ and Q for a given device geometry. Triangular cantilevers of tether angles ranging from 15 to 90 degrees are grouped together.

We find the mechanical properties of etched geometries differ significantly from drumheads. In general, the etched geometries have lower resonance frequencies and less damping. Figure 3 shows the measured frequencies and quality factors for several device geometries. The beams (with width of 1000 nm) and crosses (with cross bar widths of both 600 nm and 1000 nm) display lower $f_0$ and similar quality factors compared to drumheads. Therefore, the average damping relative to drumheads decreases by ~50% for the 600 nm cross and more modestly for the beams and 1000 nm crosses. Damping reduction is more pronounced for the cantilever geometries; for the tethered cantilever (Figure 1(c)), we observe $f_0$=10.11±1.22 MHz and Q=137.19±31.54 leading to an average damping of 7.5% that of the drumheads. For the triangular cantilever devices (Figure 1(d)), we find $f_0$=3.79±1.16 MHz, Q=467.74±166.55, and a mechanical damping coefficient that is 1.1% of that seen in the drumheads.

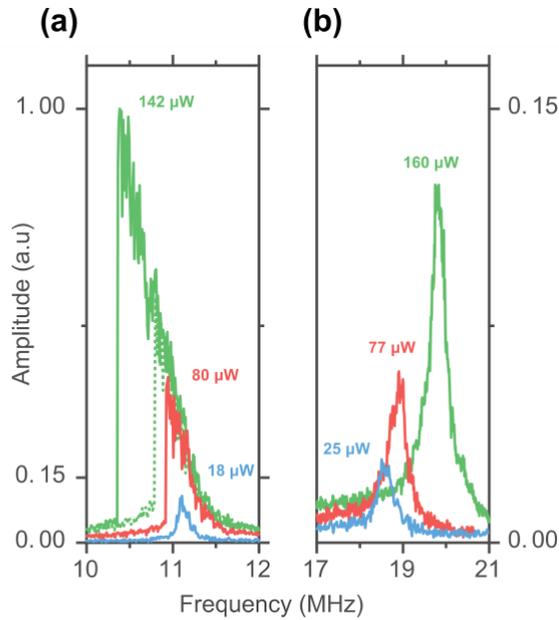

**Figure 4: Comparison of the amplitude response of a tethered cantilever and of an unmodified drumhead.** (a) Amplitude response of a tethered cantilever device at increasing optical drive powers. At high drive power, the resonance frequency lowers and the amplitude response curve become multi-valued and displays hysteresis; traces going from high to low frequency are shown solid green, while those going from low to high frequencies are shown in dashed green. (b) Amplitude response for a typical drumhead device; plotted on the same scaled y-axis as (a) and at increasing drive power. A 10-fold decrease in the amplitude response is observed for the drumhead compared to the tethered cantilever. A trend towards increasing frequency with higher optical drive powers is seen with the drumheads, likely due to thermal contraction of the graphene as it is heated by the DC component of the optical drive.

The cantilevered geometries presented here are unique due to their large aspect ratios, up to 1.66 in this work compared to less than 1 for previously fabricated graphene cantilevers[25]. We observe a significantly increased amplitude response for both types of cantilevers, roughly by a factor of 10 compared to drumheads at similar optical drive powers (Figure 4). This result is expected due to the lower resonance frequencies and larger displacements of cantilevers. We are able to use this transduction sensitivity to resolve a thermally driven resonance for triangular cantilever devices with no external drive (Figure S7), which we are unable to see for any other device geometry.

We also observe a markedly enhanced nonlinear response, typical of a softened duffing oscillator[20] for all measured tethered cantilevers at low optical drive powers of ~20-40 µW. For comparison, drumheads were driven up to ~350 µW with no discernable departure from a linear response. This can be understood by realizing that the onset of geometric non-linearity in cantilevers scales with the aspect ratio[39]. Another factor could be a large strain-driven tension in the tethers. Finite element simulations (Figure S8) on the tethered cantilever geometry show that large strain-driven tension arises in the tethers during mechanical oscillations, which could contribute to the observed non-linear behavior in this structure.

Similar non-linearities have been exploited in other nanomechanical systems to reduce noise[40,41], tune quality factors[42], couple mechanical modes[43], or as a means to improve mass sensitivity[44]. Although this type of non-linearity has been observed previously in graphene[10,15,18], the comparative drive powers reported here to achieve a large non-linear response indicates that the tethered cantilever geometry could be an ideal device architecture for future studies of non-linear graphene nanomechanics.

In terms of reducing damping, the triangular cantilever geometry proved the most promising. This geometry consists of two ~750 nm long, ~200 nm wide tethers supporting a central platform (Figure 1(d)) with the angle between the tethers ranging from 15 degrees to 120 degrees. Due to the low bending rigidity of graphene, many of the devices flip upwards to some degree (Figure

S4). There could be an additional degree of stabilization of the cantilevers due to contamination leftover from the fabrication process. A typical amplitude response curve for a device with a 90 degree tether angle is shown in Figure 5(a); this device has a Q=628. From this data, we see that the triangular cantilevers generally have frequencies 80% lower than drumheads but have higher Q, and lower mass, yielding a damping coefficient that is two orders of magnitude smaller than the value for drumheads. We observe that the mechanical damping decreases with tether angle, reaching its minimum value at 30 degrees. Our data from these measurements is summarized in Figure 5(b). A device with a 30 degree tether angle gave a measured Q of 849, which is the highest Q to date for a graphene cantilever at room temperature[25]. In this case, the damping dropped to 0.47% the value for drumheads. One key difference between the triangular cantilever and the other device geometries is that its structure cannot sustain much in-plane tension, suggesting that low stress, low tension graphene mechanical resonators may yield lower damping. Because smaller angles support less tension, we would expect them to yield higher Q, in accord with our findings.

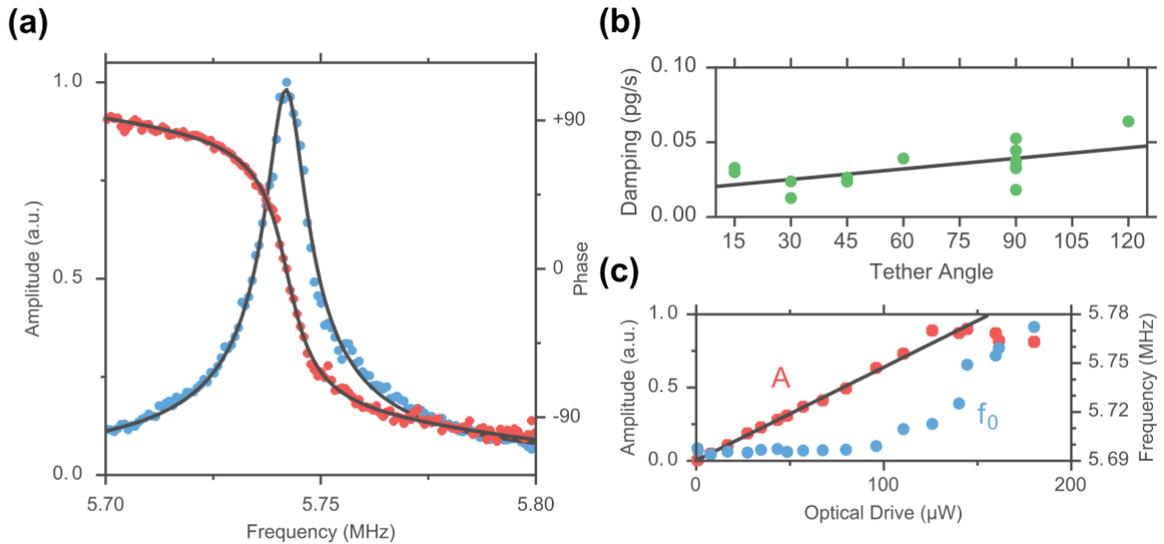

**Figure 5: Mechanical properties of a triangular cantilever.** (a) Amplitude (blue) and phase (red) response of a device with a 90 degree tether angle. The response curves are fit to a driven

damped harmonic oscillator model (black lines) with a Q=628. (b) Mechanical damping plotted against the tether angle for the triangular cantilever devices. A trend towards higher dissipation is seen as the frame angle increases. (c) Amplitude (red) and frequency (blue) as function of optical drive power. The black line is a linear fit to the amplitude response data. A linear response is observed for low drive powers. At high drive powers, a reduction in the measured amplitude and an irreversible increase to the frequency is seen, likely due to structural changes in the device.

To further explore the role of tension in triangular cantilevers, we investigated the effect of optical drive power on the amplitude and resonance frequency, both shown in Figure 5(d), for a device with a 90 degree tether angle. First, we find the amplitude increases linearly and reversibly over a ~100 µW range of optical drive, setting a minimum dynamic range of 33 dB. Furthermore, the response remains Lorentzian over the entire power range, unlike the tethered cantilevers, which go non-linear at relatively low power. Over the same power range, we find that $f_0$ remains relatively constant. The invariability of $f_0$ in the reversible regime and the broad linear response give a strong indication that any structural changes due to power absorption (*i.e.* thermal expansion or contraction, larger oscillation amplitude) do not lead to appreciable increases in tension in these devices, thereby lending validation to the claim that the triangular cantilever cannot sustain much in-plane strain. In contrast, drumheads experience large frequency shifts as the optical drive power is increased (Figure S9) due to the device and substrate heating. The insensitivity of the triangular cantilever frequency to optical drive power is attractive for force and mass sensing, since small changes to the resonator environment due to pump laser noise and other sources do not cause undesirable changes in the frequency. At higher drive powers (above 120 µW), we see irreversible changes, with $f_0$ increasing and the amplitude decreasing. Post-measurement SEM imaging reveals that devices driven past the reversible regime suffered from structural deformation such as out-of-plane buckling and kinking, leading to a shorter cantilever, smaller reflective surface area, and, consequently, the observed increase in resonance frequency and decrease in transduction sensitivity (Figure S4).

**Discussion**

The amount of pre-tension present in the shaped devices relative to the drumheads offers insight into the observed decrease in damping seen in all FIB cut geometries. Previous work[15] has identified local strain coupling to surface defects as the most likely source of damping in fully clamped graphene drumheads. Of the geometries considered here, the triangular cantilever geometry has the lowest tension and thus we would expect it to have the lowest strain-induced dissipation, consistent with our measurements. Similar investigations of low-tension[13] or minimally clamped[18] graphene mechanical resonators have also observed relatively high quality factors and low damping, which agrees with our result. Thus, strain reduction in devices could be a possible route towards high quality factor, ultra-sensitive graphene devices.

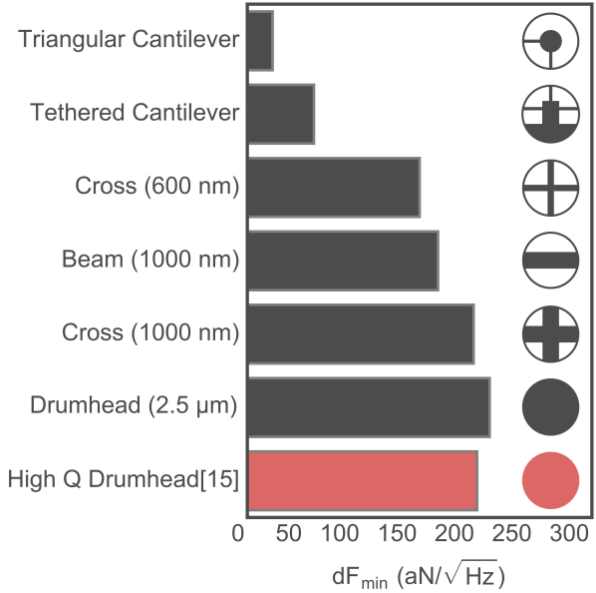

**Figure 6: Role of geometry on the minimum detectable force of graphene nanomechanical resonators.** Triangular cantilevers displayed the lowest minimum detectable force of all measured devices, with a value of ~22 aN/$\sqrt{\text{Hz}}$. For context, the characteristic force sensitivity (~200 aN/$\sqrt{\text{Hz}}$) for a high-quality factor graphene drumhead[15] is indicated in red.

The triangular cantilevers we present here operate in a unique mechanical regime characterized by small mass (0.6 fg), low frequency (several MHz), large amplitude response, and high quality factors (up to 849). This regime offers the potential for exceptional mass and force sensitivity. For example, the theoretical minimum detectable mass[45] for our most sensitive device is 30 zg, given by $\delta M_{min} \approx \frac{b}{\pi f_0} 10^{DR/20}$, where $DR$ is the minimum dynamic range and $m_{eff}$ is the effective mass, not taking additional contaminant mass introduced during the fabrication process into account[12]. This value could be significantly improved through use of a higher dynamic range optical or electronic transduction technique[8,12,18]. We estimate the minimum theoretical detectable force[6] of ~14 aN/$\sqrt{\text{Hz}}$, given by $dF_{min} = \sqrt{8\pi k_B T m_{eff} f_0 / Q}$ for the most sensitive device measured. In contrast to drumheads, which have a force sensitivity of several hundred aN/$\sqrt{\text{Hz}}$ for all sizes [15], we see a strong dependence on the device geometry. We compare the triangular cantilevers to the drumheads, beams, and crosses (Figure 6) and observe a factor of ~10 enhancement to the force sensitivity, corresponding to a reduction of ~100 in the mechanical damping. These triangular cantilevers, along with the recently reported lithographically patterned H resonator, constitute the highest reported force sensitivities for room temperature graphene mechanical resonators[18]. It is noteworthy that both of these devices utilize patterned, low-tension graphene. The FIB milling technique presented here offers an excellent method to further explore the geometric dependence of the force sensitivity, since it allows for rapid prototyping and characterization of desired device architectures.

**Conclusion**

In this work, we use FIB milling to efficiently fabricate suspended graphene structures into a wide variety of novel geometries. All shaped geometries exhibited a decrease in mechanical damping relative to the drumheads. Furthermore, we find that cantilever-style structures display additional sought-after attributes including easily accessible non-linear behavior, large transduction response, high Q, and state-of-the-art force sensitivities, while also operating in the previously inaccessible low-tension regime. Importantly, this result was achieved strictly though simple geometric shape tuning of commercial graphene, in the absence of complex fabrication techniques or ultra-clean graphene. Our findings indicate a close relationship between geometry, tension, and mechanical characteristics: structures that support less tension, such as the triangular cantilever, have lower dissipation, while structures with concentrated tension, such as the tethered cantilever, exhibit strong non-linearity. Thus, our FIB shaping technique offers a prescription to tailor key nanomechanical properties of graphene through geometry. In particular, our work gives a well-defined, repeatable approach to achieve high-Q, low-mass graphene devices. Our approach can be easily extended to shape graphene for other nanomechanical device applications, such as creating coupled mechanical resonators or phononic crystal cavities. It can also readily be applied to shape other 2D materials, such as hexagonal boron nitride and molybdenum disulfide, to explore the interplay between geometry and optical and electronic properties not present in graphene, such as photoluminescence and piezoelectricity.

**Methods**

We measured the graphene nanomechanical resonators in vacuum ($10^{-6}$ Torr) at room temperature using all-optical actuation and transduction. We mechanically drove the devices with a focused (50X, N.A. = 0.55 microscope objective), amplitude modulated 445 nm blue diode laser that we operated up to 350 μW. Transduction of device motion was accomplished using a Michelson Interferometer; in our setup, we used a 532 nm green single longitudinal mode DPSS laser, and reference-arm feedback to maintain a constant sensitivity to mechanical motion. Green power incident on the device was kept constant at 350 μW across all measurements. To address individual devices, we used a three axis motorized positioning system for lateral position and focus control of the lasers. We monitored the interferometric signal with a fast photodiode and used lock-in amplification (Zurich Instruments HFLI2) to recover the amplitude and phase response of the graphene devices.

**Acknowledgments**

The authors acknowledge facilities and staff at the Center for Advanced Materials Characterization in Oregon (CAMCOR). We would especially like to thank Joshua Razink and Jordan Mohrhardt for assistance acquiring the TEM data. We also would like to thank Andrew Blaikie, Rudy Resch, Kara Zappitelli, and Josh Ziegler for scientific discussions and review of this manuscript.

# Supporting information for "Shape tailoring to enhance and tune the properties of graphene nanomechanical resonators"

*David Miller, Benjamín Alemán*

Department of Physics, University of Oregon, Eugene, Oregon 97403

Material Science Institute, University of Oregon, Eugene, Oregon 97403

Center for Optical, Molecular, and Quantum Science, University of Oregon, Eugene, Oregon 97403

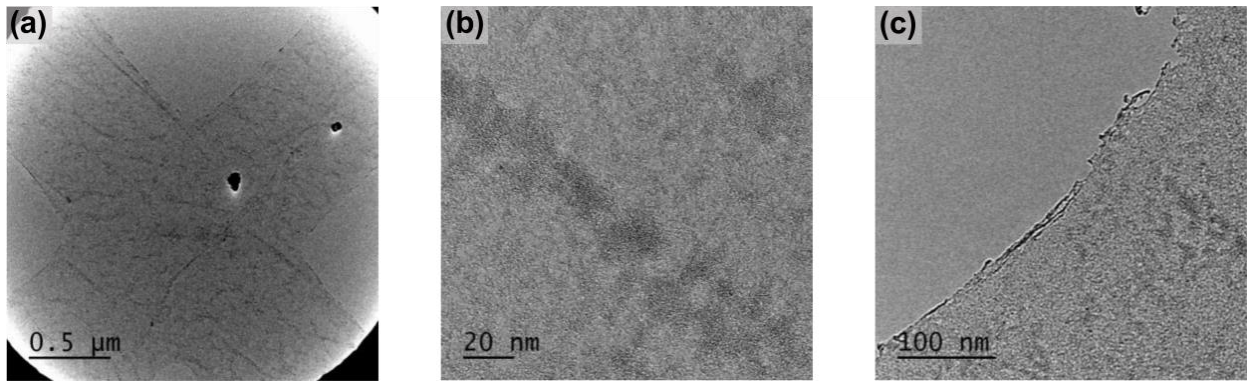

**Figure S1: Transmission Electron Microscopy (TEM) images of Focused Ion Beam (FIB) cut devices taken with an FEI Titan operating at 80kV.** (a) TEM image of a graphene cross. The black dots are contaminants remaining from the transfer process. (b) Higher magnification image of the same graphene cross. Polymer contamination leftover from the transfer process is visible as the darker contrast regions. (c) Rough edge after FIB milling similar to the one shown in the main text.

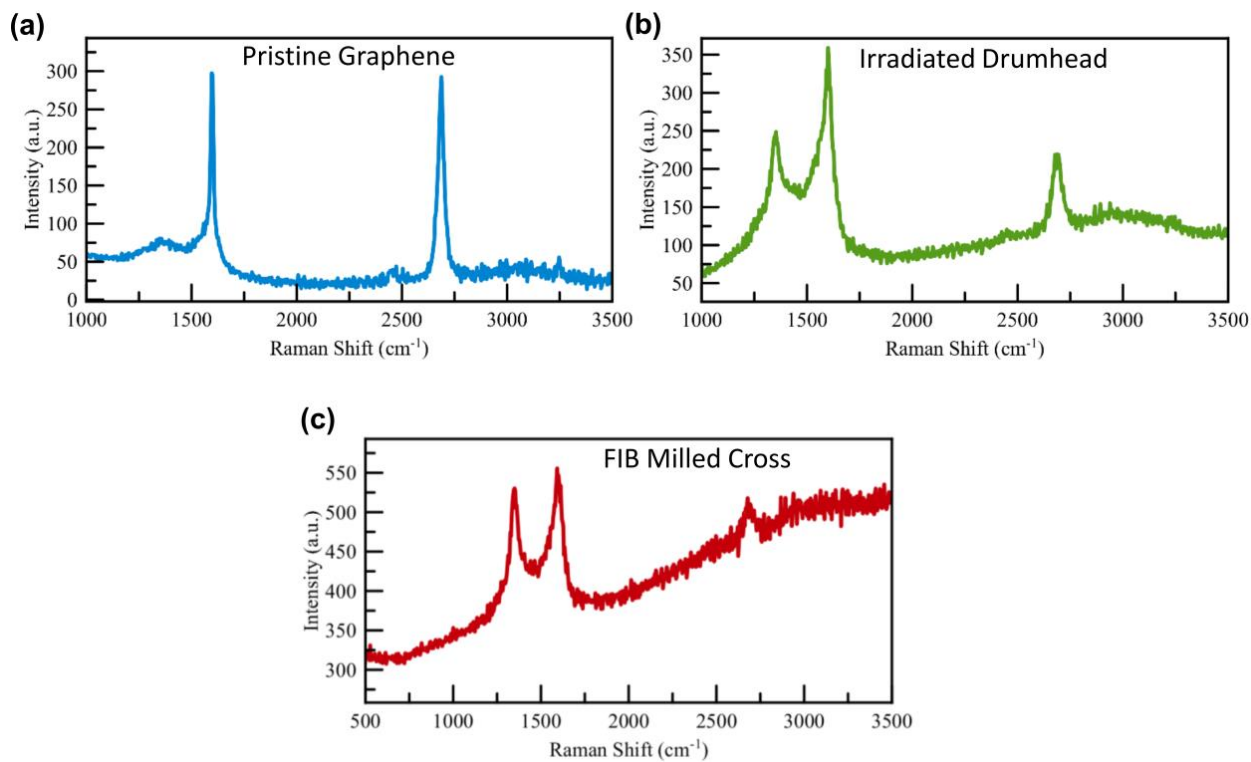

| | $I_G$ (a.u) | $I_{2D}$ (a.u) | $I_D$ (a.u) | G (cm$^{-1}$) | 2D (cm$^{-1}$) | D (cm$^{-1}$) | FWHM$_G$ (cm$^{-1}$) | FWHM$_{2D}$ (cm$^{-1}$) | FWHM$_D$ (cm$^{-1}$) |
|---|---|---|---|---|---|---|---|---|---|
| Pristine Graphene | 245±3.3 | 256.5±2.9 | 30.0±0.9 | 1598.0±0.2 | 2686.6±0.2 | 1366.9±3.6 | 7.7±0.3 | 32.2±0.5 | 94.13±5.7 |
| Irradiated Drumhead | 213.5±15.5 | 97.14±2.7 | 139.3±5.8 | 1600.8±1.2 | 2688.4±1.4 | 1355.5±6.0 | 23.3±5.8 | 46.6±1.8 | 65.58±10.0 |
| FIB Milled Cross | 129.2±3.7 | 38.70±17.7 | 178.0±2.4 | 1600.9±0.6 | 2682.0±15.7 | 1347.6±0.4 | 55.8±3.0 | 68.7±44.4 | 66.8±1.6 |

**Figure S2**: **Raman spectroscopy of FIB milled graphene**. (a) The Raman spectrum of the graphene prior to milling has an $I_{2D}/I_G$ ratio of ~1 and a sharp 2D peak with a FWHM of ~32 cm$^{-1}$. The $I_{2D}/I_G$ is low compared to as-grown CVD monolayer graphene but is typical of annealed, single-layer graphene [1], as described by the supplier of the graphene. We also fit a broad peak underneath the G peak, which is indicative of carbonization of hydrocarbon residue during annealing. (b) The nanomechanical drumheads that have been briefly exposed to the ion beam show signs of damage and modified lattice strain, evidenced by an increased D peak intensity and a lower 2D peak intensity [2]. The FWHM of both the G and D peaks increases as well. (c) The milled devices show a continuation of the trends seen in the FIB exposed drumheads, indicating a higher defect density. Additionally, the edges of the cut devices are expected to contribute to the enhanced D-peak [3]. All Raman spectra were obtained with a WITEC alpha300 Raman microscope with a 532 nm excitation laser. The laser power was kept low to avoid damaging or heating graphene.

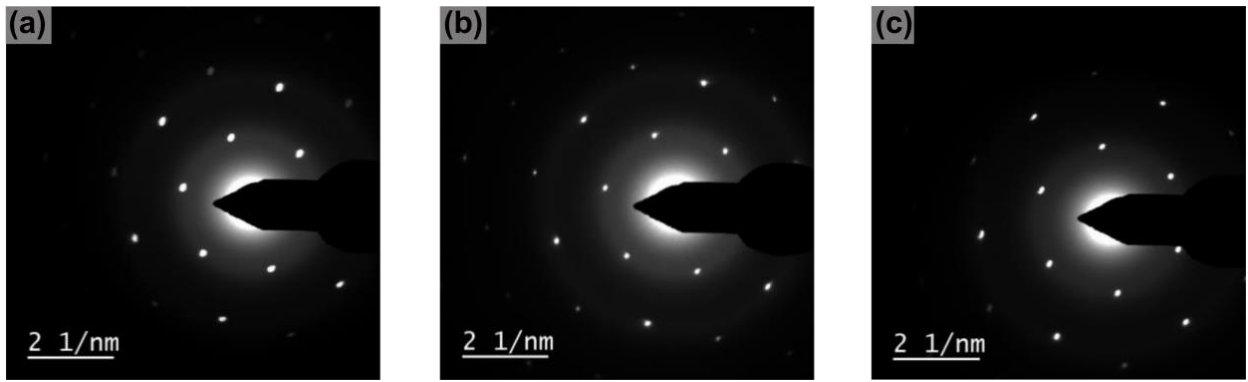

**Figure S3**: **Selected Area Electron Diffraction (SAED) of the graphene devices before and after FIB irradiation.** (a) SAED of a pristine graphene drumhead imaged far away from the milled region. The single set of diffraction spots confirms that the graphene is single grain. Some slight rotation of diffraction spots is observed and is likely due differential strain and fold defects in the graphene. (b) SAED image of a graphene drumhead which has been irradiated with a FIB 'snapshot', equivalent to a dose of ~ .0007 pC/µm². (c) SAED of a graphene cross. Despite the FIB irradiation, the graphene possesses a diffraction pattern corresponding to single-crystal graphene.

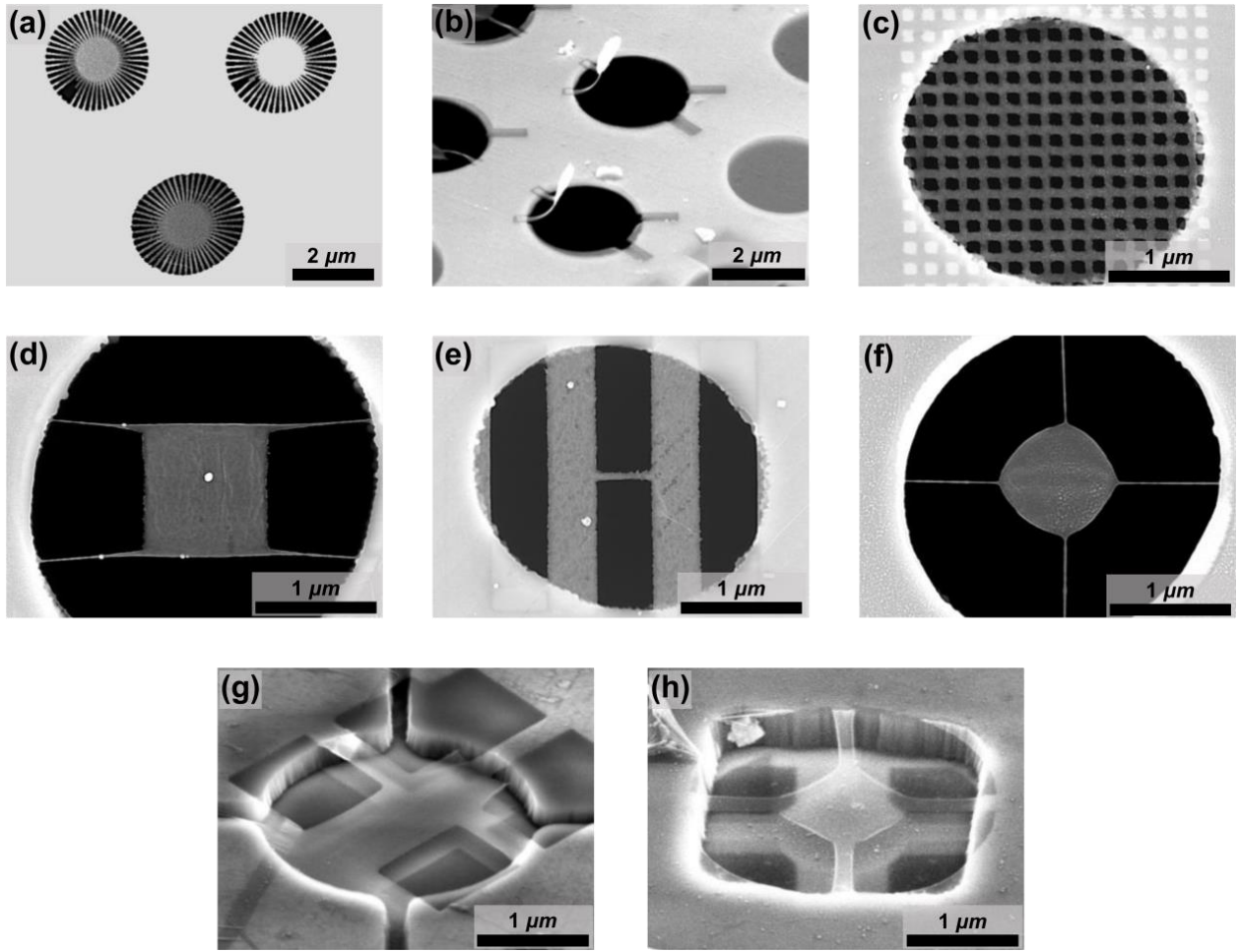

**Figure S4: Additional SEM images of selected devices** (a) "Trampoline" devices consisting of 48 individual tethers ~40 nm in width. (b) Due to the low bending rigidity of graphene, all cantilevers bend upwards to some degree but the effect is exaggerated after optical characterization at high power (c) Mesh cut into graphene with pitch ~100 nm. (d) Doubly-clamped suspended 'H' structure. The tethers are similar to the scrolled graphene shown in the main text. (e) Finished coupled beam geometry with ~500 nm beams mechanically coupled through a ~50 nm tether. (f) Trampoline style device with tethers of scrolled graphene. (g) Tethered cantilever style resonator fabricated over a cavity using CVD graphene transferred using the techniques described in [4]. (h) Trampoline resonator fabricated over a cavity.

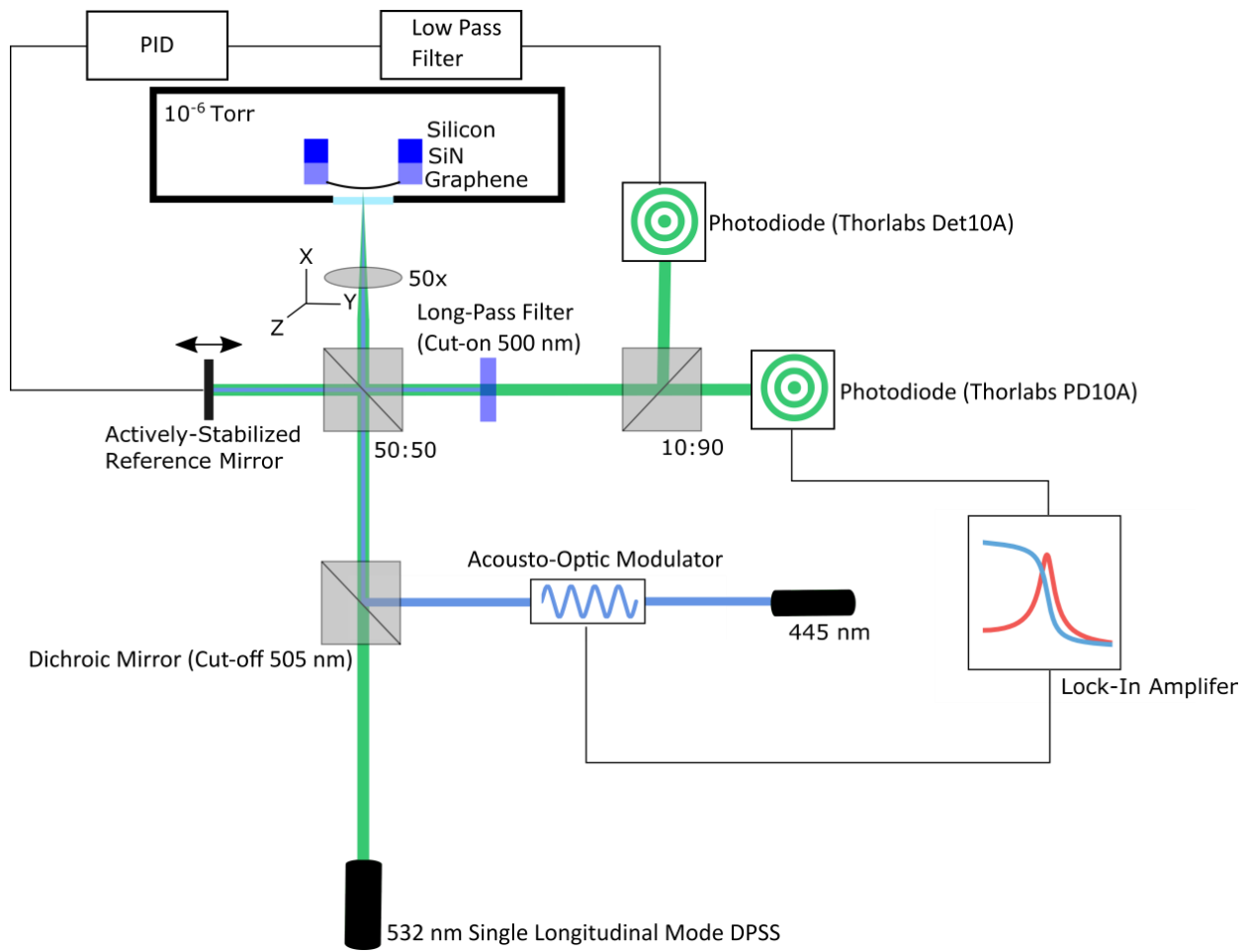

**Figure S5: Detailed diagram of the interferometric measurement of graphene mechanical motion**. An incident 532 nm single longitudinal mode laser is divided by a 50:50 beamsplitter into a signal and reference arm. Reflected light from the graphene devices and a reference mirror is interfered on two fast photodiodes using a 10:90 beamsplitter. The reflected signal is fed through a low-pass filter with a characteristic time constant much longer than the period of the mechanical resonance frequencies of the graphene devices. This filtered signal is used as the input for a PID loop. The output of the PID loop drives a piezoelectric crystal, which adjusts the length of the reference arm to compensate for low-frequency path length changes in the interferometer. The transmitted signal is measured using a lock-in amplifier to recover amplitude and phase

information. A 445 nm diode laser is amplitude modulated via an acousto-optic modulator and coupled into the optical path through a dichroic mirror in order to photothermally actuate the mechanical motion. Prior to detection, the 445 nm light is filtered out by a long-pass filter.

**S6: Calculation of strain for a mechanical drumhead.** We follow the calculation given in [5] to calculate the minimum strain in the graphene drumheads. The fundamental resonance frequency for a tensioned membrane is given by $f_0 = \frac{4.808}{2\pi D}\sqrt{\frac{Yt\epsilon}{\rho\alpha}}$, where D is the drumhead diameter (2.5 μm), t is the thickness (.335 nm), Yt is the in-plane Young's modulus (340 N/m), $\rho$ is the two-dimensional mass density ($7.4\times10^{-16}$ g/μm$^2$), α is scaling factor to account for additional contaminant mass from the transfer process, and $\epsilon$ is the strain in the membrane. Since $\alpha$ is unknown, typically of order 1, we set a minimum, rather than absolute, value on the strain. Using the measured resonance frequency of 21.54 MHz for the drumheads, we calculate a minimum strain of $\epsilon \sim 10^{-5}$, which is in accord with previous measurements of graphene drumheads on holey silicon-nitride in [5].

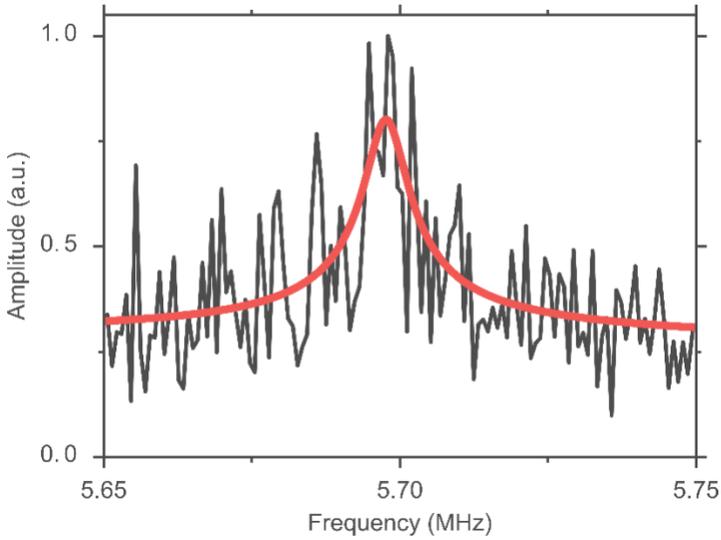

**Figure S7: Thermomechanical noise (black) for a triangular cantilever fitted to damped harmonic oscillator fit (red).** The time constant on the lock-in amplifier was set to 10 ms.

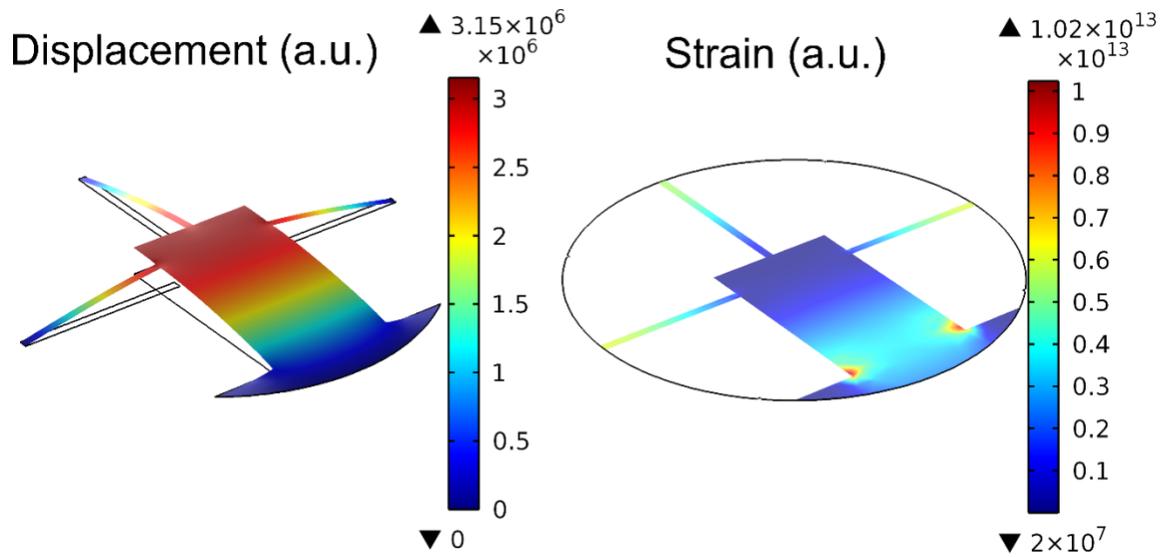

**Figure S8: Finite element simulations of normalized displacement and strain for tethered cantilever.** Regions of high strain are visible both in the base of the cantilever and at the ends of the three tethers.

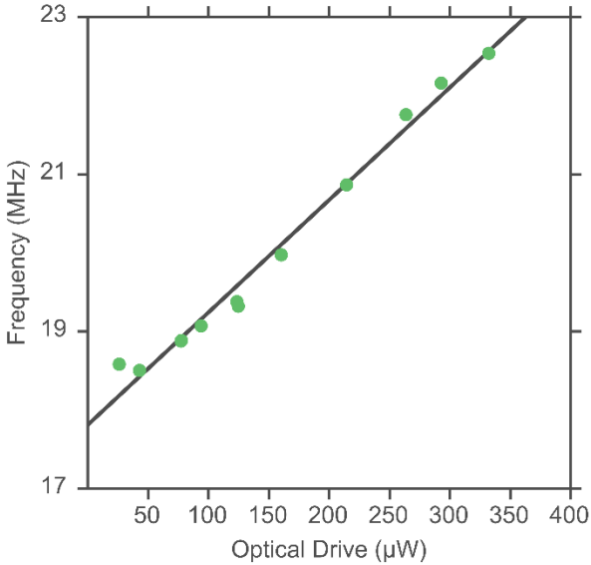

**Figure S9: Frequency shift as a function of optical drive power for the drumhead shown in Figure 4(b)**. The frequency is seen to increase with increasing optical drive power. We attribute this to a combination of thermal expansion of the silicon-nitride and thermal contraction of the graphene at increased temperatures[6,7]. The amplitude response remains linear over the entire range of optical drive power. A linear fit (black line) shows a frequency shift of 14 kHz/µW over a range of about 4 MHz.

**Supporting Videos SV1-SV3:** *In-situ* SEM videos of FIB milling for several geometries. Each of these videos was taken over the course of about 1 minute.

**Video SV1:** Failure of a graphene cross during FIB milling from asymmetric tension due to incorrect cutting methods

**Video SV2:** Successful fabrication of a graphene cross, achieved by first outlining the shape of the device with a series of vector cuts.

**Video SV3:** Fabrication double clamped beam through single pass raster cutting.